%% file: nature-template.tex
\documentclass{nature}


\pdfoutput=1
\usepackage{amssymb,amsmath,epsfig,times,color,hyperref}

\title{The split in the ancient cold front in the Perseus cluster}

\author{Stephen A. Walker$^{1}$\thanks{Email: 
    stephen.a.walker@nasa.gov} , John ZuHone$^{2}$, Andy Fabian$^{3}$, Jeremy
Sanders$^{4}$}

\begin{document}

\maketitle

\begin{affiliations}
 \item Astrophysics Science Division, X-ray Astrophysics Laboratory, Code 662,
NASA Goddard Space Flight Center, Greenbelt, MD 20771, USA
 \item Harvard-Smithsonian Center for Astrophysics, 60 Garden St., Cambridge, MA
02138, USA
 \item Institute of Astronomy, Madingley Road, Cambridge CB3 0HA
 \item Max-Planck-Institute fur extraterrestrische Physik, 85748 Garching,
Germany
\end{affiliations}

\begin{abstract}
Sloshing cold fronts in clusters, produced as the dense cluster core
moves around in the cluster potential in response to in-falling subgroups,
provide a powerful probe of the physics of the intracluster medium (ICM), and
the magnetic fields permeating it\cite{Markevitch2007,Zuhone2016review}. These
sharp discontinuities in density and
temperature rise gradually outwards with age in a characteristic spiral pattern,
embedding into the intracluster medium a record of the minor merging activity of
clusters: the further from the cluster centre a cold front is, the older it is.
Recently it has been discovered that these cold fronts can survive out to
extremely large radii in the Perseus cluster\cite{Simionescu2012}.  Here we
report on high spatial
resolution Chandra observations of the large scale cold front in Perseus. We
find that rather than broadening through diffusion, the cold front remains
extremely sharp (consistent with abrupt jumps in density) but instead is split
into two sharp edges. These results show that magnetic draping can suppress
diffusion for vast periods of time, around $\sim$5 Gyr, even as the cold front
expands out to nearly half the cluster virial radius.
\end{abstract}

One of Chandra's first advances in understanding galaxy clusters was the
discovery of cold fronts: sharp edges in X-ray surface brightness where the gas
density rapidly drops while the temperature rises, the opposite of a
shock\cite{Markevitch2000,Vikhlinin2001}. The exceptional sharpness of these
cold fronts, smaller than the Coulomb mean free path, indicates that diffusion
and conduction processes in the ICM (which should smooth the edges in density
and temperature) and hydrodynamic instabilities are suppressed, most likely due
to magnetic draping\cite{Lyutikov2006,Asai2007,Dursi2008,ZuHone2011}. Detailed
studies of sloshing cold fronts have at present
been limited to those in the central $\sim$100kpc, where the ICM is brightest
and the
cold fronts are young. 

One remarkable recent development is the discovery of large scale cold fronts
reaching out to very large radii, around half the virial radius (730kpc) in
Perseus\cite{Simionescu2012} and several other
clusters\cite{Rossetti2013,Walker2014}. These colossal cold fronts are the
oldest coherent structures surviving in cool core clusters, with the one we
present here in Perseus being around $\sim$5 Gyr old (see Figs 1, 2, and
supplementary figure 3). By comparison the structures seen in cluster cores
produced by AGN feedback are being constantly dissipated and replaced on
timescales of just $\sim$100s Myr. 

As cold fronts rise outwards the resulting velocity shear should lead to
Kelvin-Helmholtz instabilities (KHI)\cite{ZuHone2010,Roediger2012,Roediger2013},
as we have found in younger cold
fronts\cite{Walker2017}. Magnetic draping over the cold front surfaces has been
found to
inhibit KHI\cite{Walker2017,Werner2016}. In ancient cold fronts, much more time
has elapsed over which
KHI can develop, and over which diffusion and conduction processes can smooth
over the jumps in density and temperature, providing powerful tests of
simulations\cite{ZuHone2011,ZuHone2018}. 

These ancient cold fronts have risen slowly outwards from the core, experiencing
vastly different ICM environments. In cluster cores AGN feedback
dominates\cite{Fabian2012},
with sound waves\cite{Fabian2017} and turbulent gas motions\cite{Zhuravleva2014}
likely dissipating the feedback
energy into the ICM. They then move away from the core, where AGN feedback no
longer dominates. Moving into the cluster outskirts, the turbulence and gas
motions in the ICM should rise again\cite{Lau2009}, due to the continuing
accretion of
matter. By studying Chandra observations of the ancient cold front in Perseus,
we can see the time integrated effect of all of these processes on the cold
front. 

The existing shallow XMM mosaic of Perseus is shown in Fig. 1 (top). Overlaid on
this is the gradient map from Gaussian Gradient Magnitude
filtering\cite{Sanders2016b,Walker2016} of the
inner Chandra and outer XMM mosaics, emphasising the edges. The large cold front
730kpc to the east of the core is clear. From these wide-field data we obtain a
large scale, low spatial resolution temperature map (Fig. 1, bottom left),
showing the characteristic cold arch of gas behind the surface brightness edge.
This is consistent with the expected temperature behaviour seen in
simulations\cite{ZuHone2011,ZuHone2018}, (Fig. 1, bottom right), for sloshing
perpendicular to the line
of sight.  

\begin{figure}
  \begin{center}

\includegraphics[width=0.9\linewidth]{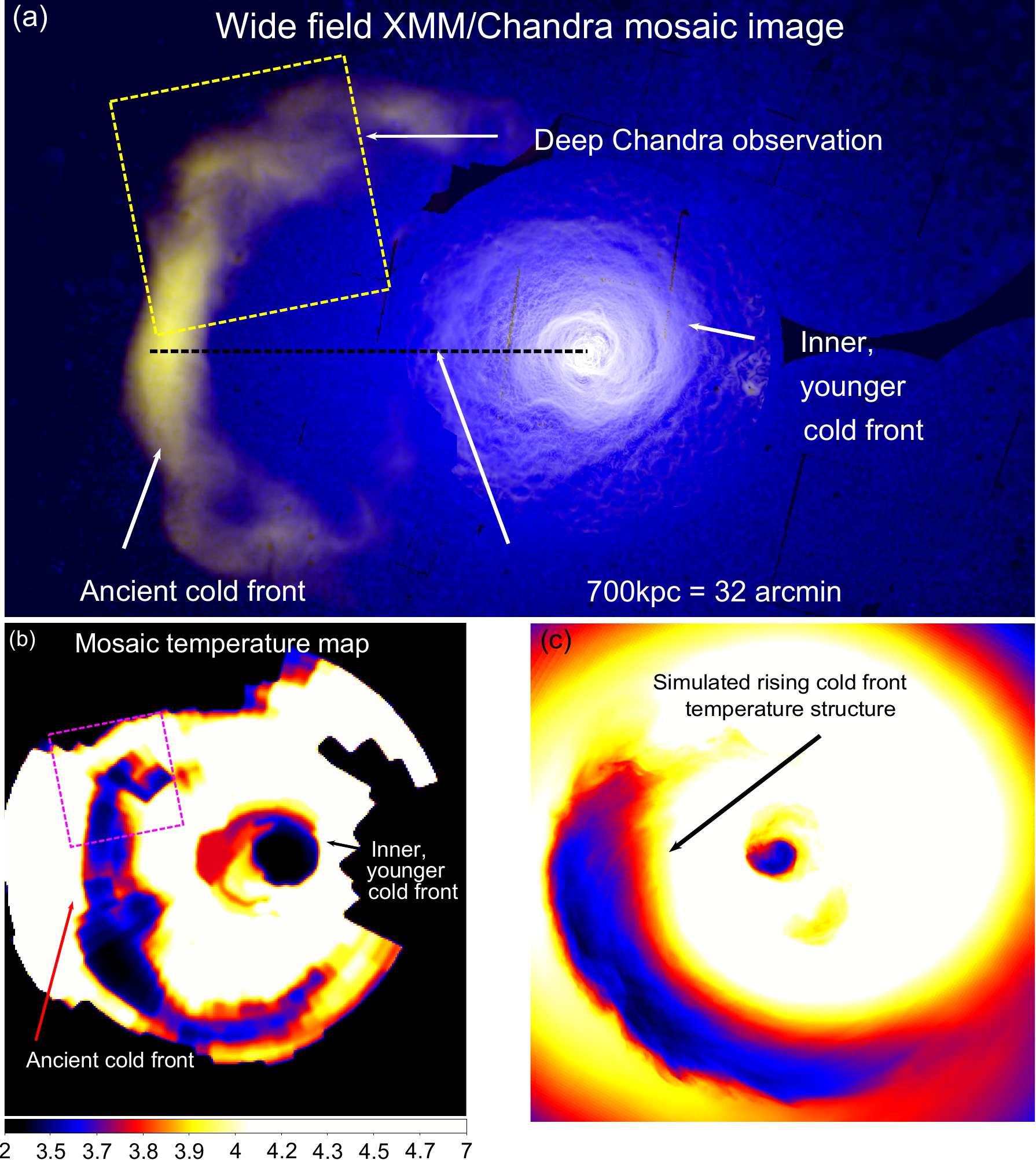}
  \end{center}
\vspace{-1.0cm}
\caption{\textbf{Wide-field X-ray observations compared to simulations.}  \emph{Top:}
XMM-Newton
mosaic of the Perseus cluster (blue), over which is laid the gradient map
(white, pink and yellow) obtained from GGM filtering of the Chandra (inner) and
XMM (outer) mosaics. The ancient, 730kpc radius outer cold front is clear on the
left. The location of our new deep Chandra observations (presented in Fig. 2) is
shown by the dashed box. \emph{Bottom left:} Low spatial resolution large scale
temperature map from the shallow wide-field XMM and Chandra mosaics, which shows
the large cold arch of gas behind the ancient cold front. \emph{Bottom right:}
Temperature map from gas sloshing simulation\cite{ZuHone2018}. The arch shaped
cold band
matches the geometry of the observed cold front arch seen in Perseus in the
bottom left panel.}
\end{figure}

Our new, high spatial resolution Chandra ACIS-I observations (Fig 2, left) have
targeted the region to the north of the large scale cold front, (Fig. 1, dashed
square). These show that, instead of the typical single cold front edge in the
X-ray surface brightness, the cold front has split into 2 edges, separated by
$\sim$100kpc. The temperature map (Fig 2, right) shows that there are two cold
arcs
behind each X-ray surface brightness edge, separated by a band of hotter gas.
Such cold front splitting has never been seen before.

The two surface brightness edges are remarkably sharp. When the surface
brightness edges are fit with  two broken powerlaws smoothed by a Gaussian
(Supplementary figure 1), their widths are consistent with being zero. The
conservative (3$\sigma$) upper limit on the width of both edges is 8 arcseconds
(3.1
kpc). This is much smaller than the Coulomb mean free path for diffusion from
the inner (brighter) side to the outer side of  $\lambda_{\rm {in \rightarrow out,2}}$
=23.2$\pm$6 kpc (60.0$\pm$15
arcsec) for edge 2, and $\lambda_{\rm {in \rightarrow out,1}}$ =9.6$\pm$1.5 kpc
(25$\pm$4 arcsec) for edge 1. This
shows that magnetic fields are still able to support the edges against
broadening  even at these extreme radii, and for huge periods of time. 

Comparing our observed structures to simulations of sloshing with different
strength initial cluster magnetic fields\cite{ZuHone2011,ZuHone2018} we find
that one possibility for
this `double edge' cold front splitting in this region may be due to the
development of features similar to the onset of a Rayleigh-Taylor (RT)
instability (Fig. 3 and 4). In an RT instability, dense, low-entropy gas is
suspended ``above" (in the sense of the direction of the local gravitational
acceleration) less dense, higher-entropy gas. This suspension is unstable, and
produces the familiar RT ``fingers'' of dense gas falling into the lower-density
material.

In these simulations, the cooler gas is initially displaced from the dark matter
dominated potential by the ``ram-pressure slingshot'' effect\cite{Ascasibar2006}
(Fig 3, top left).
As the gas falls back into the potential and overshoots it, it forms the first
cold front. The lowest-entropy gas behind this front falls back into the cluster
potential minimum (Fig. 3, top right) from a location slightly offset from the
center of the front (similar to the RT ``fingers''), while the rest of the front
continues to propagate radially outward. The result is that the coldest,
lowest-entropy part of the sloshing structure develops a ``bend' from the initial
front going counterclockwise inward toward the cluster center, but protruding
outward from the edge of this bend is a ``hook'' feature which is a continuation
of the curvature of the original front (Fig. 3, bottom). These simulations begin
with an initial uniform thermal to magnetic pressure ratio
($\beta=p_{\rm {th}}/p_{\rm {B}}$) throughout
the cluster before any sloshing begins. As the sloshing proceeds, the magnetic
field becomes amplified and wrapped around the cold fronts, affecting the
development of the ``hook'' feature as we show later in Fig. 4.  In our best
matching simulation, the magnetic field draped along the cold front is typically
10 times stronger than the field outside the cold front, and is $\sim$5$\mu$G in the
oldest simulation time slice. 

\begin{figure}
  \begin{center}
\includegraphics[width=\linewidth]{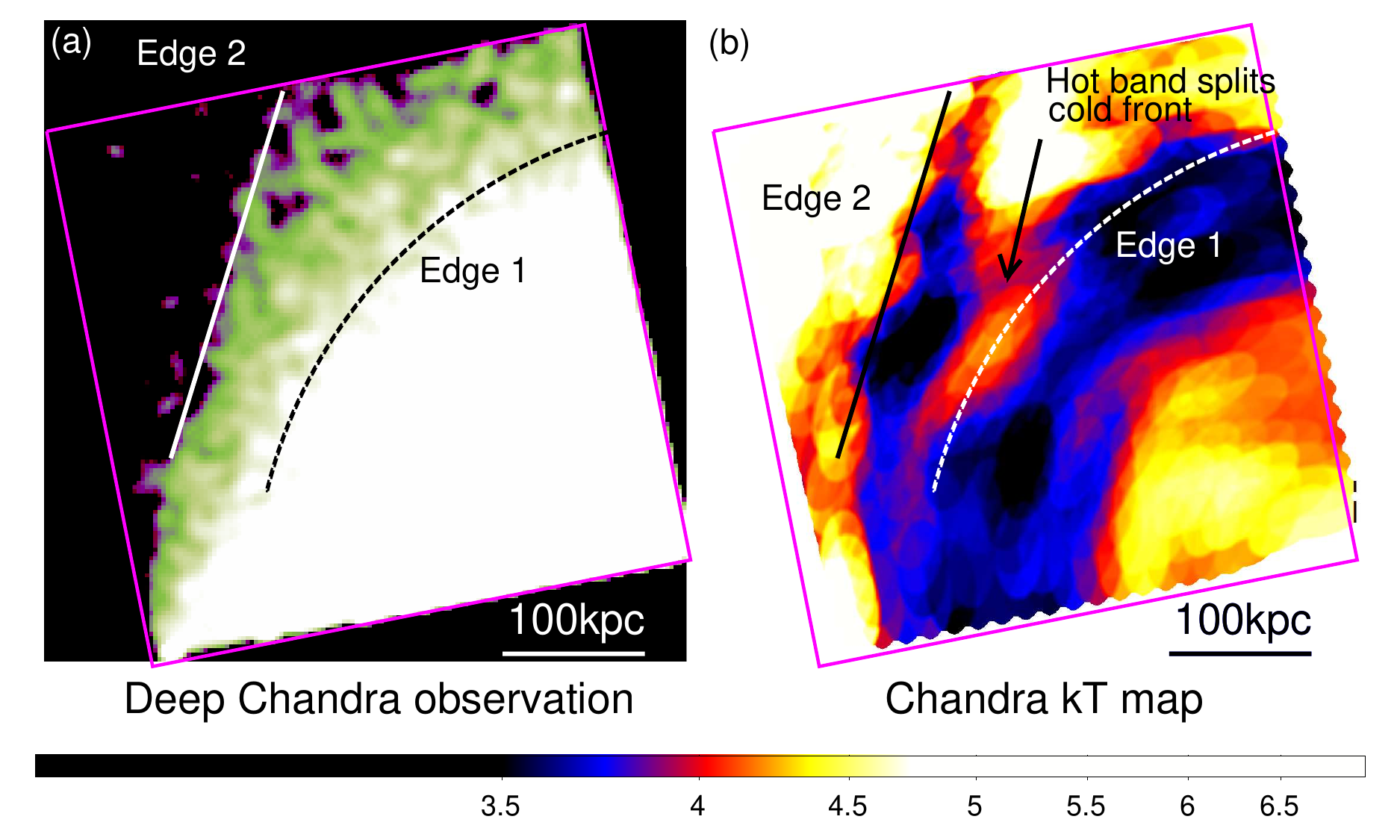}
  \end{center}
\vspace{-1.0cm}
\caption{\textbf{Deep Chandra observations.} The left panel shows an image in
the 1.2-4.0keV band from
our 95 ks Chandra observation, which maximises the contrast of the edges. Two
edges are visible in the deep observation: `edge 1' follows the curvature of the
cold front shown in Fig 1, while the outer `edge 2' is approximately straight.
The right hand panel shows the temperature map, showing a sharp temperature jump
at each edge, and a hot band splitting the cold front. The colorbar shows the
temperature in keV. }
\end{figure}

This hook feature rises outwards with the cold front, and survives out to the
final epoch of the simulation, at which the cold front is 3.3Gyr old (we define
the age as the time since closest approach of the interacting clusters) and has
risen to a radius of 450kpc (supplementary figure 4).  Plotting the cold front
radius against age from the simulation (supplementary figure 3), the relation is
perfectly linear (a constant rise speed, as has been found in earlier
works\cite{Roediger2012}).
Since the thermodynamic and mass profiles of the cluster continue as simple
powerlaws in the simulation from 100kpc outwards, it is  possible that the hook
can survive out to the observed cold front radius of 730kpc, with the cold front
continuing to rise with constant speed. Extrapolating the linear best fit, the
Perseus cold front at 730kpc would be 5.2Gyr old, which we conservatively quote
throughout as $\sim$5Gyr. 

This hook feature leads to two edges forming, separated by a hotter band, which
continues to move outwards as the cold front rises, similar to what we observe
(Fig. 4, middle column for the $\beta$=200 simulation). When the magnetic field
is
stronger (initial $\beta$=100), this splitting is suppressed, and the hot band
separating the two edges does not form (Fig. 4, left column). When the magnetic
field is weaker (initial $\beta$=500), the magnetic field is too weak to stop
the cold
front from breaking up into KHI, so the split cold front structure does not form
(Fig. 4, right column). 

While the simulations we use here only extend to a cold front age of 3.3Gyr,
previous simulations\cite{Ascasibar2006,ZuHone2010} have successfully produced
cold fronts out to 4.7Gyr
from the time of closest approach, comparable to the $\sim$5Gyr estimated here
for
Perseus. We note that our simulations do assume the rigid potential
approximation\cite{Roediger2012b}, whose validity is unclear outside 500kpc,
(though simulations
which do not use the rigid potential approximation\cite{Ascasibar2006}, also
show the hook feature
persisting to large radii and later epochs, so its presence and maintenance is
not an artifact of this approximation). We stress that we are proposing a
general explanation for the cold front splitting we observe, and attempting to
reproduce it exactly in simulations is beyond the scope of this paper. 

The double edge feature we observe may also be compatible with multiple edges
seen in simulations of the Virgo cluster cold front for low (0.1$\%$ Spitzer)
viscosity\cite{Roediger2013} due to KH rolls, though at present these have only been simulated on
much smaller scales ($\sim$10s kpc). The splitting may be compatible with a
range of
Reynolds numbers as well, but determining this is beyond the scope of this work.

The remarkable sharpness of the cold front edges shows that the magnetic field
is able to dominate over  the increasing levels of gas motions and turbulence
expected in the outer regions of clusters\cite{Lau2009}. Over the last $\sim$5 Gyr Perseus
has
continued to grow as gas accretes onto it from the surrounding cosmic web in a
highly asymmetric manner, yet this appears to have had no impact on the cold
front structure. It also shows that many generations of rising bubbles from AGN
feedback (which rise $\sim$20 times faster than sloshing cold fronts and so
should
overtake or pass through them), have had no discernible impact, suggesting that
rising bubbles can pass through cold fronts without affecting them.  

If the hook interpretation is correct, the opening angle between the two edges
is sensitive to the angle between the line-of-sight and the normal to the plane
of the sloshing, $\theta$. The opening angle between the two edges increases as
$\theta$
increases (supplementary figure 5). The observed opening angle in our Chandra
observation is $\sim$30 degrees, from which we infer that $\theta \leq$20
degrees (i.e. the
sloshing is mostly perpendicular to the line of sight, which is also in
agreement with the observed spiral structure). Recently it has been found that
if sloshing motions dominate the line-of-sight velocity gradient seen in the
core with Hitomi\cite{Hitomi2016}, the observed line widths and shifts are
consistent with a
significant inclination relative to the line-of-sight, potentially up to
$\sim$45
degrees\cite{ZuHone2018b}.

Since the sloshing needs to be mostly perpendicular to the line of sight to
allow the hook to be seen, this may have prevented hook features from being
observed in other cold front systems. Other factors which may also have
facilitated the observation of such a feature in Perseus include the larger
spatial scale of the hook in a large scale cold front (the edges are split by
$\sim$100kpc), combined with reduced projection effects away from the cluster
core,
and the large angular size of Perseus owing to its closeness. 

\begin{figure}
  \begin{center}
\includegraphics[width=\linewidth]{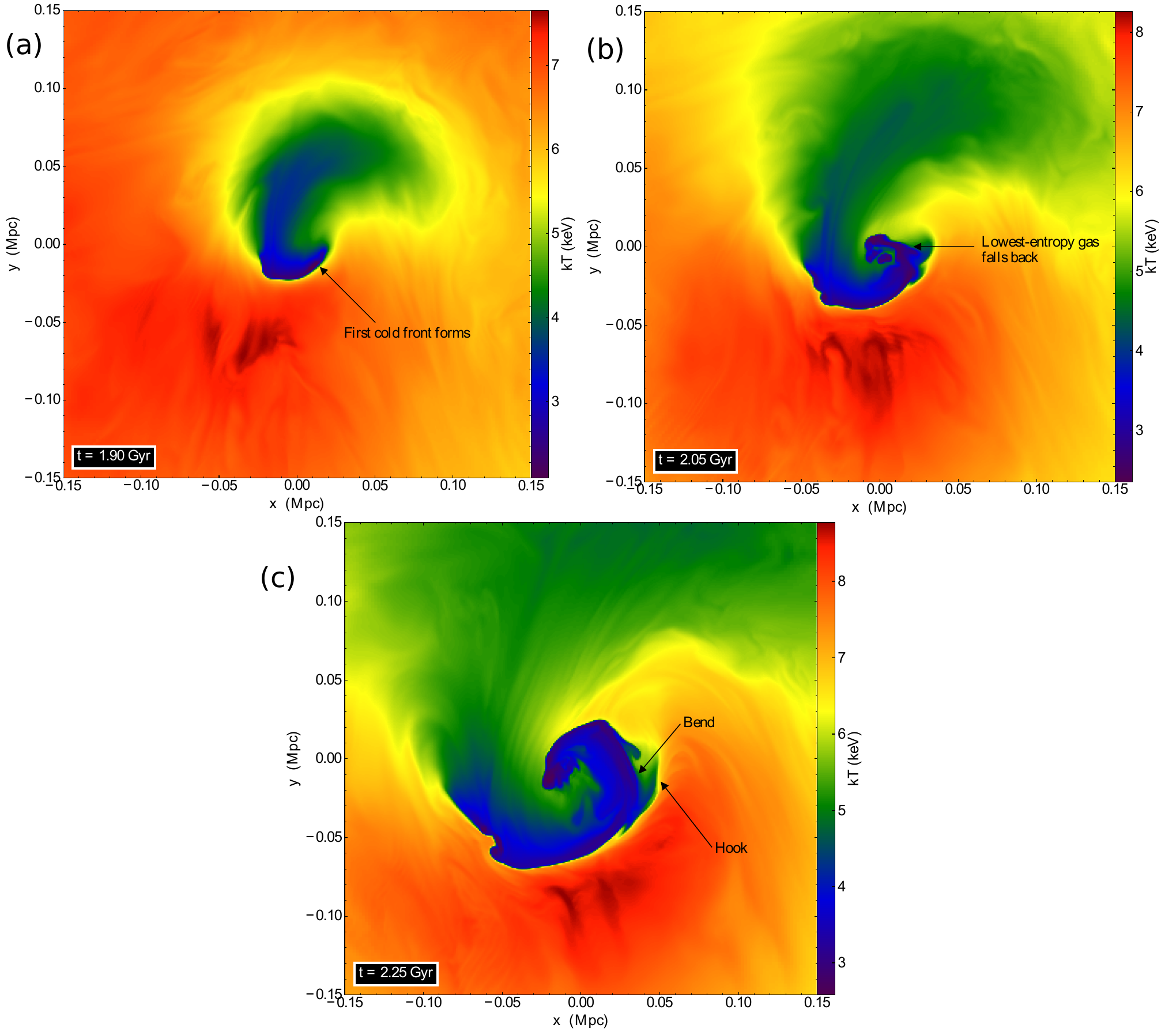}
  \end{center}
\vspace{-1.0cm}
\caption{\textbf{Simulations of cold front development with time.} Simulated
slices of the temperature maps for three time intervals showing the development
of the ‘hook’ structure as the cold front forms\cite{ZuHone2011,ZuHone2018}, as
described in the main text. The video of this simulation is available here
(\href{https://vimeo.com/236813999}{https://vimeo.com/236813999}). The time
stamp shows the time since the start of the simulation. The sloshing starts 1.3
Gyr into the simulation, so the ages of the cold fronts shown in these panels
are 0.6, 0.75 and 0.95 Gyr. The hook feature rises out with the cold front and
survives out to the final epoch covered by the simulation, at which the cold
front has risen to a radius of 450kpc (see supplementary figure 4).  }
\end{figure}

\begin{figure}
  \begin{center}
\includegraphics[width=\linewidth]{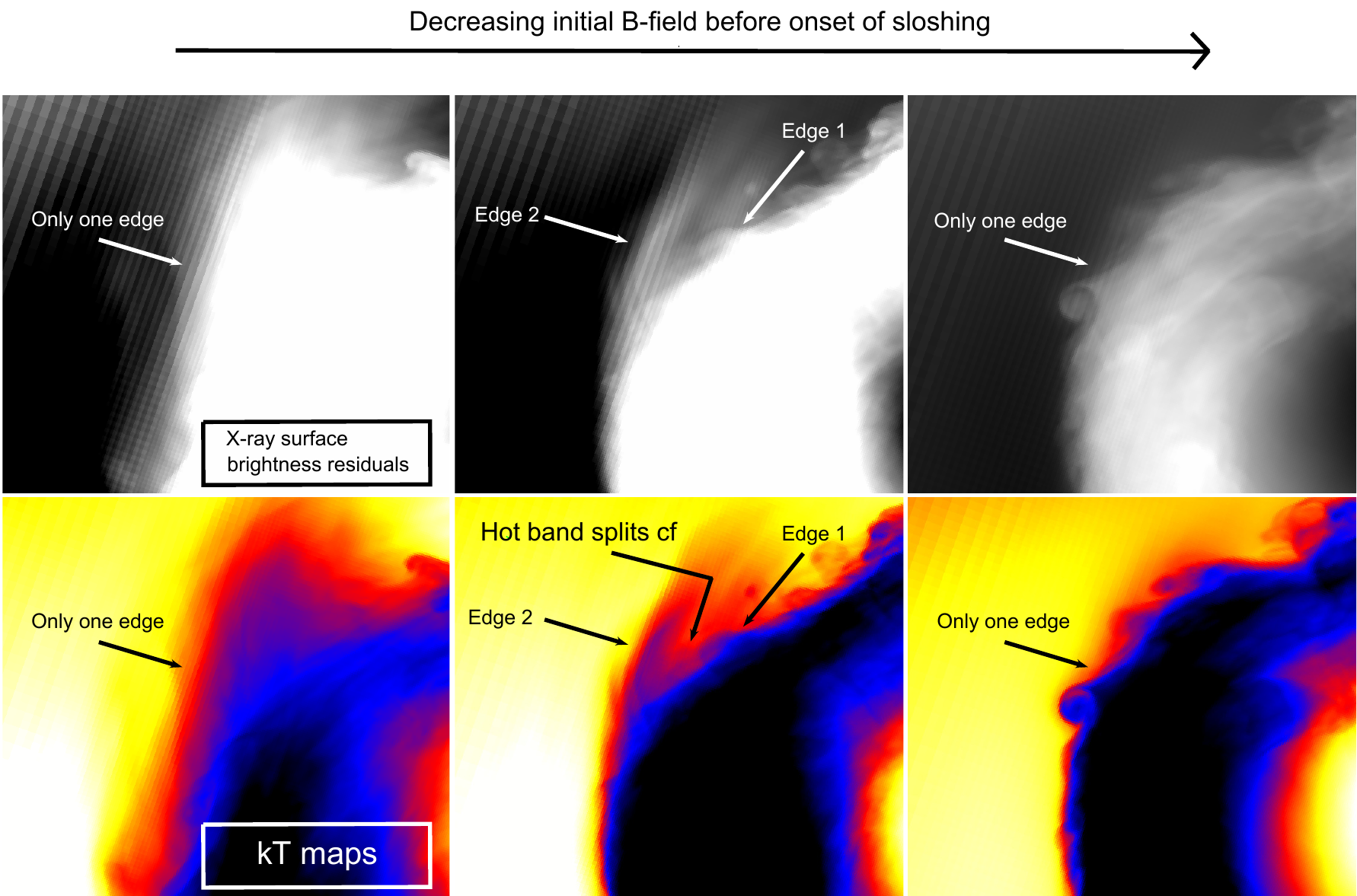}
  \end{center}
\vspace{-1.0cm}
\caption{\textbf{Simulations of cold front development with time.} Comparing
simulations\cite{ZuHone2011,ZuHone2018} of cold front development for decreasing
initial magnetic field strength from left to right. The top row shows the X-ray
surface brightness residuals after dividing out the azimuthal average. The
bottom row shows the temperature maps. In the left hand column, the initial
magnetic field is strongest ($\beta$=100) and prevents the cold front from
splitting. In the middle column, the initial magnetic field is halved
($\beta$=200), and is not strong enough to prevent the splitting, resulting in
the same double edge feature that we observe. In the right hand column
($\beta$=500),  this even weaker magnetic field cannot form the split cold
front, which instead forms KHI roll structures curving the other way. In these
time slices of the simulation, the cold front is 2.7Gyr old, but the same
behaviour with magnetic field is seen in all the epochs of the simulation.  }
\end{figure}

\subsection{References}
\bibliography{nature-template}

\subsection{Correspondence}
Correspondence and requests for materials should be addressed to S.A.W.\\
(stephen.a.walker@nasa.gov).

\subsection{Acknowledgements} 
S.A.W. was supported by an appointment to the NASA Postdoctoral Program at the
Goddard Space Flight Center, administered by the Universities Space Research
Association through a contract with NASA. A.C.F. acknowledges support from ERC
Advanced Grant FEEDBACK.

\subsection{Competing interests}
The authors declare no competing financial interests.

\subsection{Author contributions}
S.A.W. wrote the manuscript with comments from all the authors. S.A.W. performed the Chandra and XMM-Newton data analysis and lead the
Chandra proposal. J.Z. produced the galaxy cluster sloshing simulations. 

\newpage
\noindent\textbf{Methods} \\
\noindent Here we outline our analysis procedures. All errors quoted are at the
1$\sigma$ level unless otherwise specified. 

\noindent \textbf{Chandra image analysis}.  All the Chandra observations used
are tabulated in supplementary table 1. For all Chandra analysis we used the
latest version of CIAO (4.9).  CHANDRA\_REPRO was used to process the events
files, while light curves in the 0.5-7.0keV band were assessed, with the routine
LC\_SIGMA\_CLIP used to identify and remove and periods where the count rate
deviated by more than 2$\sigma$ from the mean. Background images were obtained
from stowed background, which were cleaned and processed in the same way as the
observations, and projected into the same coordinate system. These backgrounds
were rescaled to ensure they have the same high energy (9.5-12.0keV) count rates
as the observations. 

FLUX\_OBS was used to obtain background subtracted images and exposure maps in
seven narrow energy bands (1.2-1.5, 1.5-2.0, 2.0-2.5, 2.5-2.75, 2.75-3.0,
3.0-3.5, 3.5-4.0 keV), which were exposure corrected and then added, to produce
the image shown in the left hand panel of Fig. 2. WAVDETECT was used to identify
and remove point sources. 

\noindent \textbf{XMM image analysis.} The XMM observations used to produce the
0.7-7.0keV image in the top panel of Fig. 1 are tabulated in supplementary table
2. We used only the MOS observations, following\cite{Simionescu2012}. The data
reduction and image processing were performed using the XMM-Newton Extended
Source Analysis Software (XMM-ESAS), following the standard
methodology\cite{Snowden2008}. 

\noindent \textbf{Spectral analysis.} The Chandra spectral fitting and
background modelling follows the procedures described in\cite{WangWalker2014}.
Stowed backgrounds, scaled to match the 9-12 keV count rate, were used to remove
the particle background. To model the soft X-ray background, we used the best
fitting background model from the Suzaku and ROSAT background
fields\cite{Urban2014}, which consists of three thermal components modelling the
local hot bubble, the Galactic halo and a potential 0.6 keV foreground
component.  We used a powerlaw of index 1.4 to model the cosmic X-ray background
from unresolved point sources, and the normalization of this powerlaw was
calculated using the model for the cumulative flux distribution of point
sources\cite{Moretti2003}, using the threshold flux to which we can resolve
point sources in our Chandra fields, following the methods outlined
in\cite{Walker2013_Centaurus}. 

For the XMM spectral fitting and modeling, we follow the procedures
from\cite{Walker2013_CentaurusXMM}. In addition to the background model, two
gaussians were used to model the instrumental lines (Al K$\alpha$, 1.49 keV amnd
Cu K$\alpha$, 1.75 keV), with their normalisations set as free parameters.
Residual soft proton contamination was modelled as a broken powerlaw. The
powerlaw normalisation for the cosmic X-ray background from unresolved point
sources was calculated using the XMM-ESAS task POINT\_SOURCE. 

To test the robustness of the soft X-ray background modeling, we repeated each
spectral fit using the range of background component values found
in\cite{Urban2014} for the 8 different background Suzaku fields spaced equally
around the cluster. Varying the background components by these ranges had
negligible effect on the spectral fits to the intracluster medium. 

In the spectral fits we fit the intracluster medium with an absorbed
APEC\cite{Smith2001} model in XSPEC 12.9.1, using the extended C-statistic. The
column density was fixed the LAB\cite{LABsurvey} value of 1.45$\times10^{21}$
cm$^{-2}$, the metal abundance was fixed to 0.3 times solar metallicity, the
redshift was fixed to z = 0.01756\cite{Hitomi2016}, and we fit for the temperature and
density. 

\noindent \textbf{Temperature map.} The temperature map shown in Fig. 2 was
obtained by extracting spectra from a grid of partially overlapping ellipses,
which are orientated to follow the broad surface brightness distribution. Each
ellipse is shifted relative to the nearest one by half the length of its minor
axis, so that each ellipse overlaps with at least 8 neighbouring regions, over
which an average is found. The best fit temperature for each ellipse is found,
and the temperature map is produced by averaging over the ellipses. This
overlapping method acts to smooth the temperature map and reduce noise.
Extracting regions from a fixed grid ensures that we avoid any artifacts from
the binning process.  Each ellipsoidal region contains at least 2000 counts
after background subtraction.  

The widescale temperature map shown in Fig. 1 was obtained from the shallow
XMM-mosaic using larger overlapping regions consisting of elliptical sectors
spanning a radial range of 4 arcmins and with an azimuthal opening angle of 25
degrees, where the curvature of the ellipse is set to match the curvature of the
cold front's X-ray surface brightness. Each sector was shifted relative to its
neighbouring one by either a radial shift of 2 arcmins, or an azimuthal shift of
10 degrees. Again this overlapping method helps smooth the temperature map,
reducing noise. 

\noindent \textbf{Temperature profiles.} Temperature and surface brightness
profiles over directions of interest are shown in supplementary figure 1.
Profile 1 passes over the outermost edge. Profile 2 goes over the cold hook from
one side to the other.  Profile 3 goes over the hot band splitting the edges.
Profile 4 goes over the innermost edge. The temperature profiles obtained in 
supplementary figure 1 were produced by extracting spectra in regions large
enough to contain 2000 background subtracted counts, which were aligned with the
surface brightness features. The regions were spaced 8 arcsecs apart, much
smaller than the typical bin width of $\sim$0.5arcmin, to ensure that the
profile shapes were not affected by our choice of where to place our bins. The
spectral extraction regions are shown in supplementary figure 2. Deprojections
were performed using the deprojection code
DSDEPROJ\cite{Sanders2007,Russell2008}. 

\noindent \textbf{Surface brightness profile fitting and edge width measurements.} From the
deprojected temperature and density profiles across the outer edge (edge 2 in
Fig. 2 and profile 1 in supplementary figure 1), we find that the temperature
jumps from 3.75$\pm$0.5 keV to 7.0$\pm$1.0keV, while the gas density falls from
5.2$\pm$0.2$\times10^{-4}$ cm$^{-3}$ to 3.7$\pm$0.3$\times10^{-4}$ cm$^{-3}$.
The Coulomb mean free path given by\cite{Markevitch2007}

$\lambda = 15 \:{\rm kpc} \:(T/7 \:{\rm keV})^2 (n_{\rm {e}}/10^{-3} \:{\rm cm}^{-3})^{-1}    
    $        
            
\noindent inside the edge is $\lambda_{\rm {in}}$=8.0$\pm$2.1 kpc (20.7$\pm$5.6
arcsec), and outside is $\lambda_{\rm {out}}$=40.5$\pm$12.0 kpc (105$\pm$31 arcsec). 
Therefore the mean free path for diffusion from the inner (brighter) side to the
outer side, given by, 

$\lambda_{\rm {in \rightarrow out}} =  \lambda_{\rm {out}}  \quad    T_{\rm {in}}/T_{\rm {out}}   \quad
G(1)/G((T_{\rm {in}}/T_{\rm {out}})^{1/2})    $     

\noindent where
$G(x)=[{\rm erf}(x) - x {\rm erf}'(x)]/2x^{2}$  (where ${\rm erf}(x)$ is the error function) is
$\lambda_{\rm {in \rightarrow out}}$ =23.2$\pm$6 kpc (60.0$\pm$15 arcsec). From
fitting the surface brightness jump with a broken powerlaw smoothed by a
Gaussian\cite{Werner2016,Sanders2016}, the observed width of the edge is
consistent with zero, with a conservative (3$\sigma$) upper limit on the width of 8
arcsecs (3.1 kpc), which is 3.1/23.2=0.13$\pm$0.04 times smaller than
$\lambda_{\rm {in \rightarrow out}}$. The reduced $\chi^{2}$ for the best fit is 1.05
for 36 degrees of freedom. The density jump ratio is a factor of 0.7$\pm$0.06.

The pressure inside the outer edge is 2.0$\pm$0.4$\times$10$^{-3}$ keV
cm$^{-3}$, and outside is 2.3$\pm$0.5$\times$10$^{-3}$ keV cm$^{-3}$, so the
pressure is consistent with being continuous. This is not inconsistent with the
potential presence of a magnetic draping layer, since it would probably be very
thin\cite{Lyutikov2006,ZuHone2011}, and cannot be detected with the sensitivity
available in our observations.
  
For the inner edge (edge 1 in Fig. 2, profile 4 in supplementary figure 1), the
temperature jumps from 3.5$\pm$0.3keV to 5.0$\pm$0.3keV, while the density drops
from 7.0$\pm$0.2$\times$10$^{-4}$ cm$^{-3}$ to 5.7$\pm$0.2$\times$10$^{-4}$
cm$^{-3}$, which yields  $\lambda_{\rm {in}}$=5.3$\pm$1kpc (14$\pm$2.4 arcsec), 
$\lambda_{\rm {out}}$=13.4$\pm$1.7 kpc (35$\pm$4.4 arcsec) , and  $\lambda_{\rm {in
\rightarrow out}}$=9.6$\pm$1.5kpc (25$\pm$4 arcsec). Again the 3$\sigma$ upper
limit on the width of the edge is 8 arcsecs (3.1 kpc), which is 3.1/9.6=
0.3$\pm$0.05 times smaller than $\lambda_{\rm {in \rightarrow out}}$. The reduced
$\chi^{2}$ for the best fit is 1.06 for 43 degrees of freedom. The density jump
ratio is a factor of 0.8$\pm$0.05.

With the depth of data available, we are able to identify any surface brightness
jumps corresponding to density changes greater than 5 percent with at least
5$\sigma$ signficance, so any density jumps we do not identify must be less than
this 5 percent level. 

\noindent \textbf{Details of the simulations.} The simulations we used are based
on those from\cite{ZuHone2011}, but with an improved  treatment of gravity, and improved
spatial resolution\cite{Roediger2012}. In these simulations, sloshing is
initiated in a massive cool core cluster similar to Perseus
(M$_{200}$=10$^{15}$ solar mass) by an infalling smaller subcluster (one fifth
the mass of the main cluster), moving on a trajectory that gives it an impact
parameter of 500kpc, and starting from a distance of 3Mpc. The temperature, gas
density and mass profiles of the main cluster were originally designed to
emulate those observed in the similarly massive cool core cluster Abell 2029,
and have been found to provide a reasonable approximation to the Perseus
cluster\cite{ZuHone2018b}. These simulations use the rigid potential
approximation\cite{Roediger2012}, in which rather than using the N-body method
to follow the evolution of the dark matter haloes of the clusters directly, a
rigid potential is assumed for both clusters. The simulations are run in the
rest frame of the most massive cluster and corrections are applied to account
for the fact that the rest frame of the main cluster is not an inertial frame. 

\noindent \textbf{Data availability}
The data that support the plots within this paper and other findings of this
study are available from the corresponding author upon reasonable request.

\noindent \textbf{Supplementary figures}
\begin{figure}
  \begin{center}
\includegraphics[width=\linewidth]{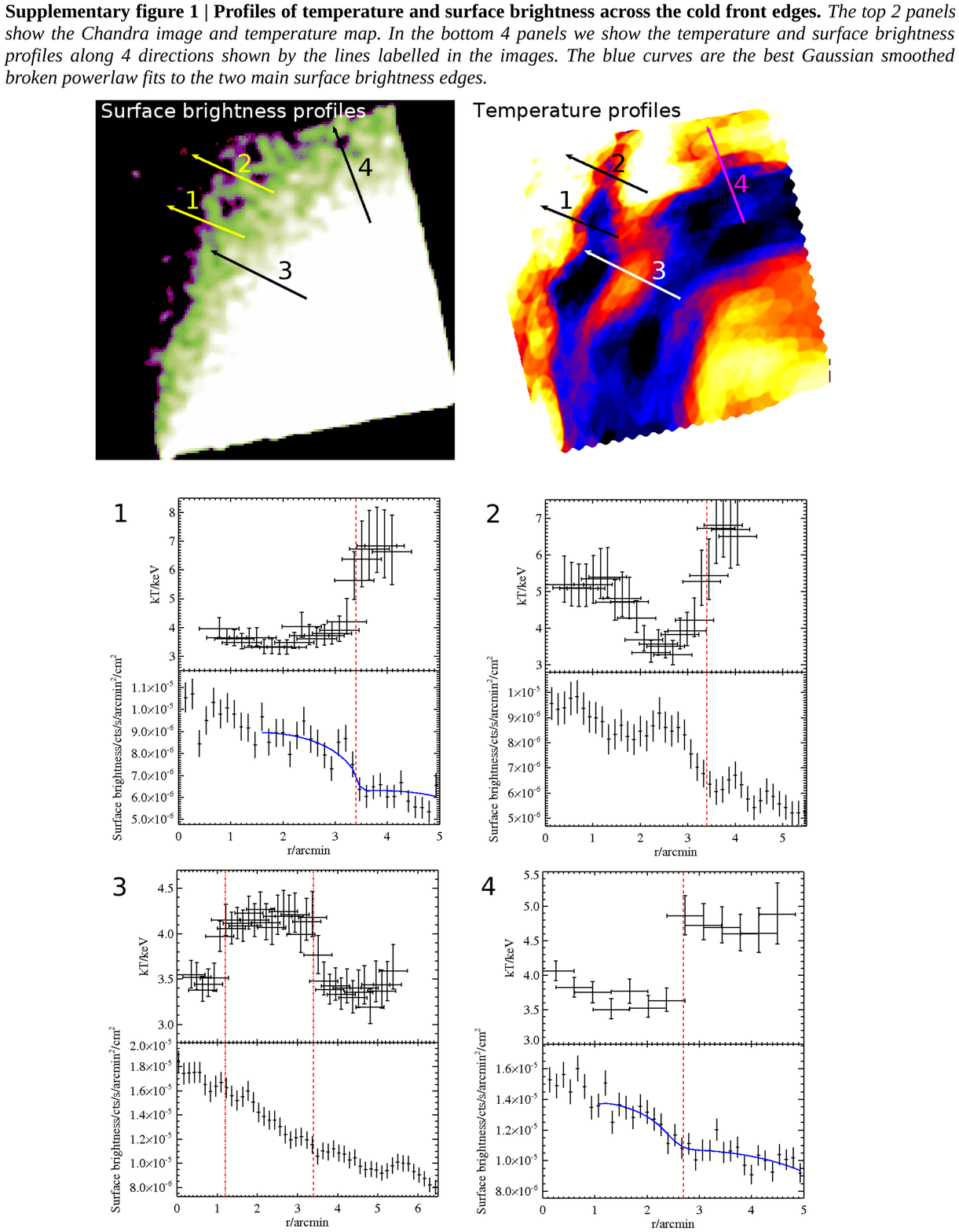}
  \end{center}
\end{figure}

\begin{figure}
  \begin{center}
\includegraphics[width=\linewidth]{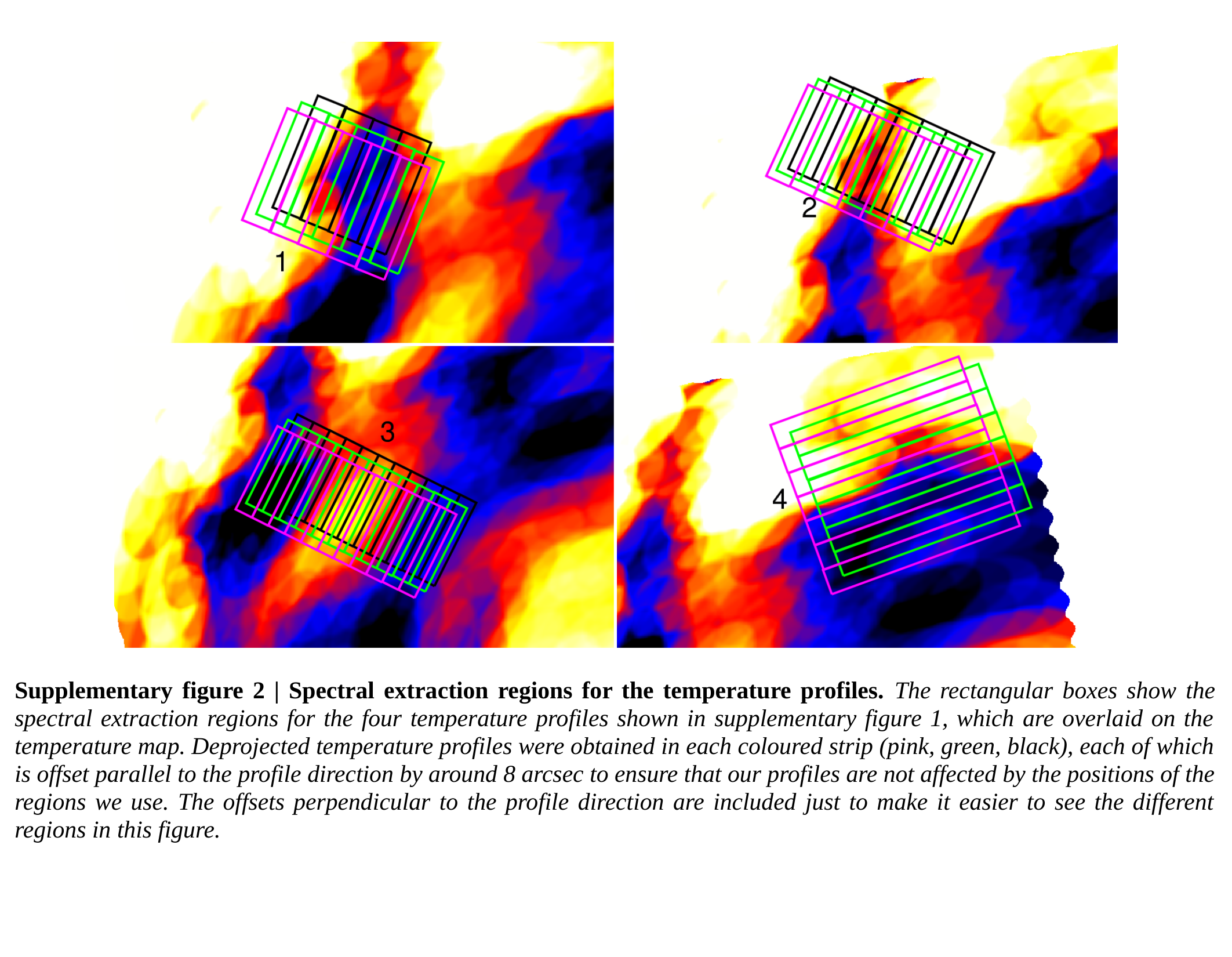}
  \end{center}
\end{figure}

\begin{figure}
  \begin{center}
\includegraphics[width=\linewidth]{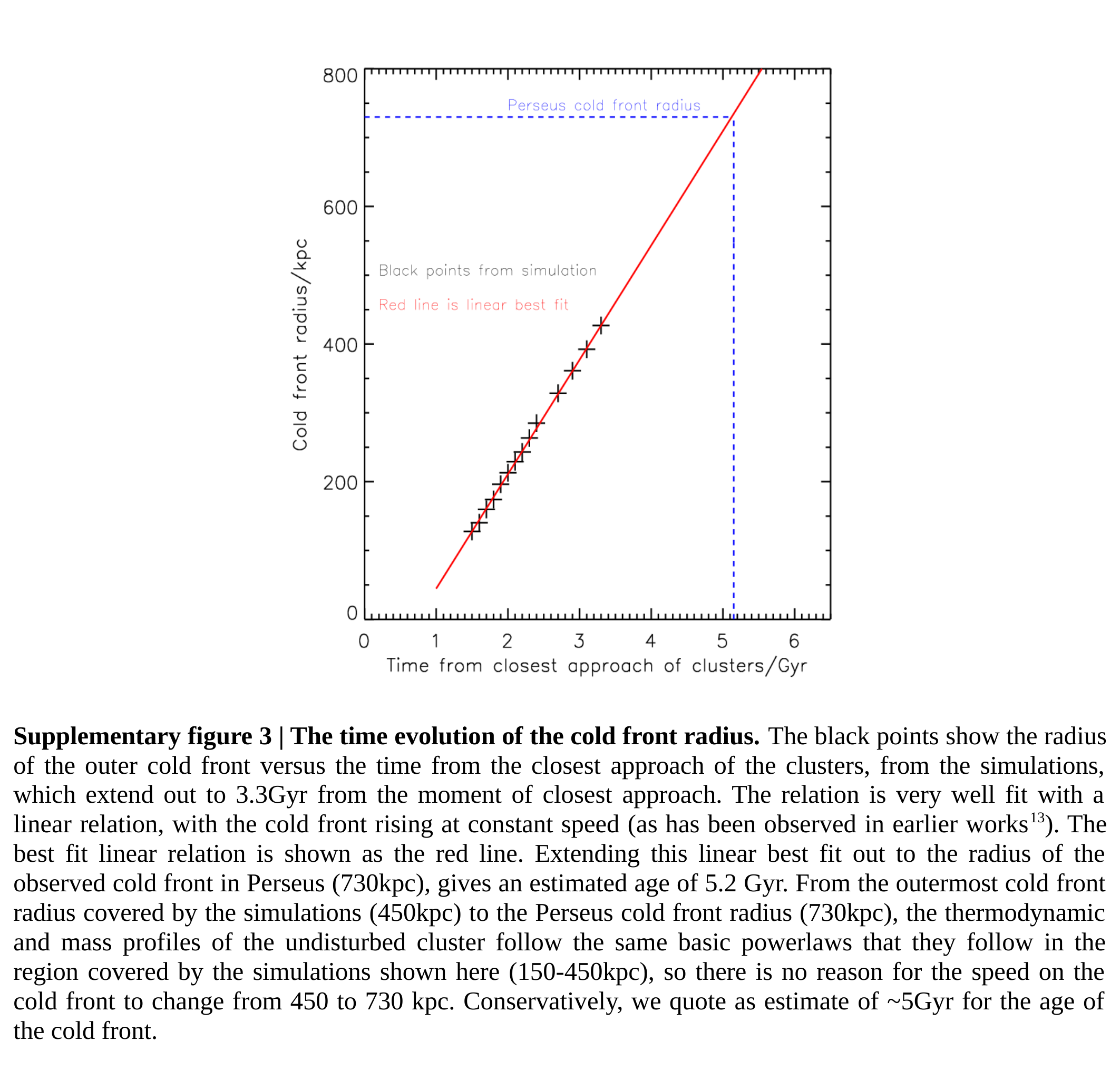}
  \end{center}
\end{figure}

\begin{figure}
  \begin{center}
\includegraphics[width=\linewidth]{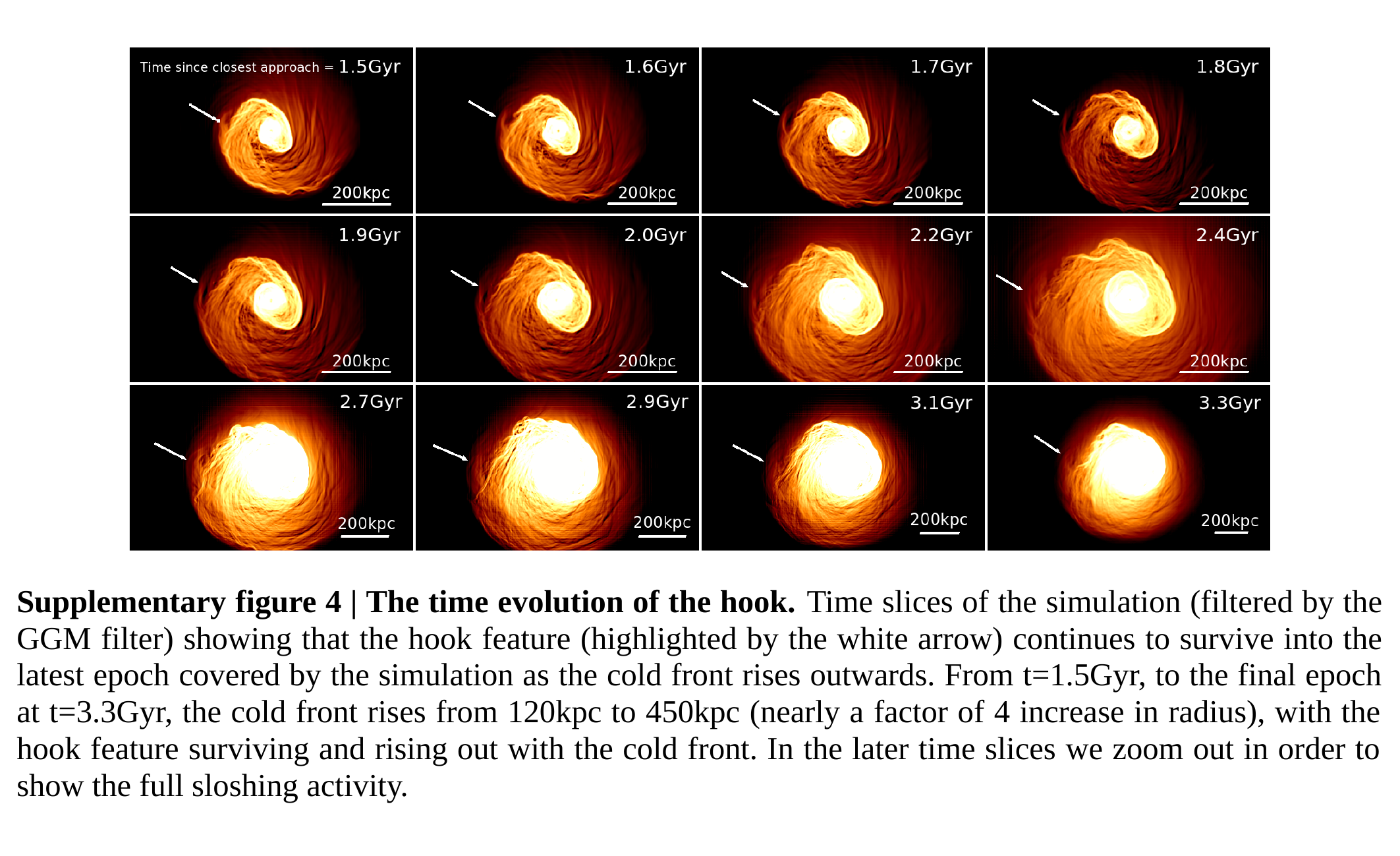}
  \end{center}
\end{figure}

\begin{figure}
  \begin{center}
\includegraphics[width=\linewidth]{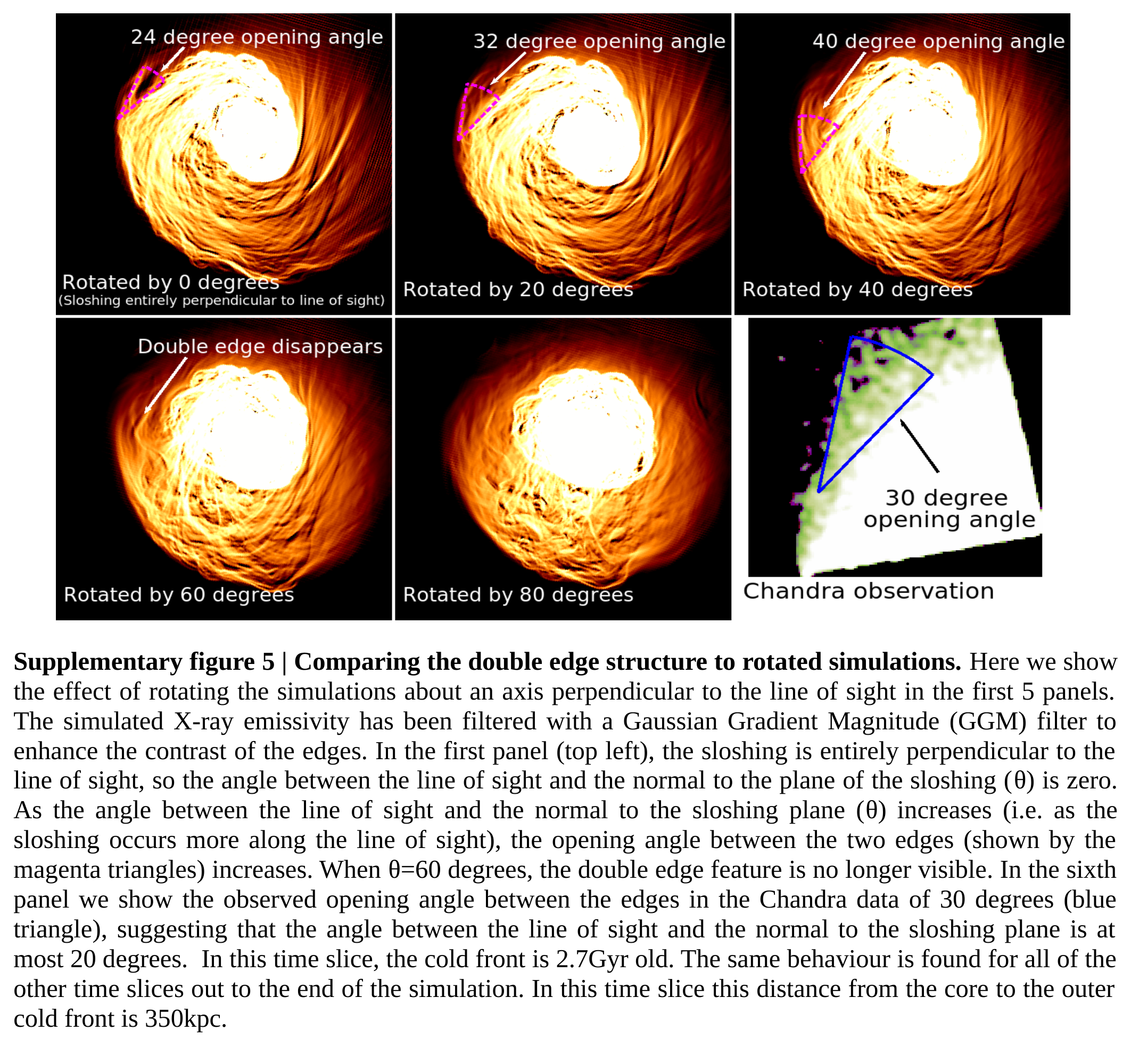}
  \end{center}
\end{figure}

\begin{figure}
  \begin{center}
\includegraphics[width=\linewidth]{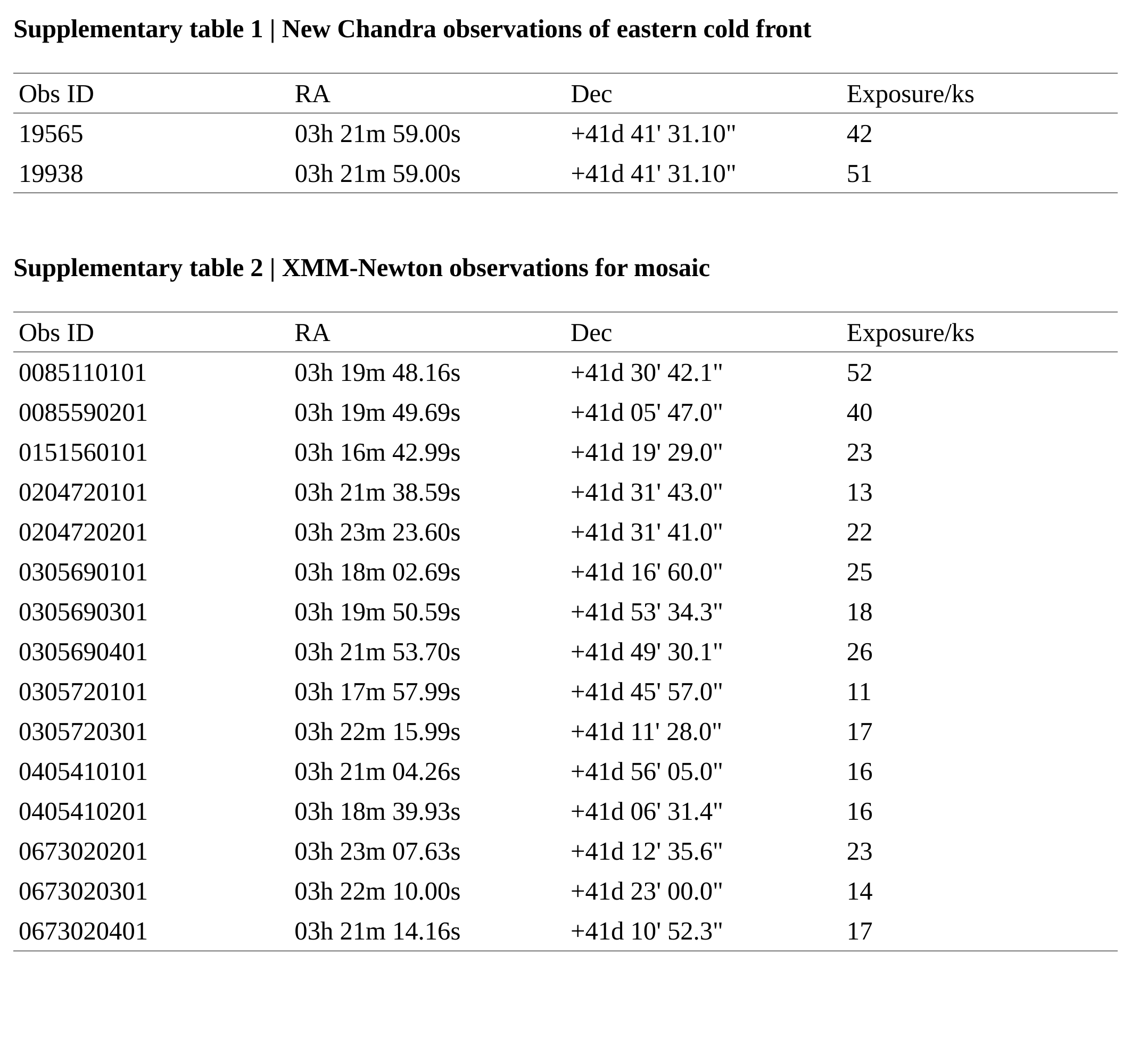}
  \end{center}
\end{figure}
%%
%% TABLES
%%
%% If there are any tables, put them here.
%%

\end{document}